\documentclass[journal]{vgtc}                     % final (journal style)
\onlineid{0}

\vgtccategory{Research}

\vgtcpapertype{application/design study}

\title{Virtual Reality Sickness Reduces Attention\\ During Immersive Experiences}

\author{%
  \authororcid{Katherine J. Mimnaugh}{0000-0002-6306-1674},
  Evan G. Center, Markku Suomalainen, Israel Becerra, Eliezer Lozano, \\ Rafael Murrieta-Cid, Timo Ojala, Steven M. LaValle, and Kara D. Federmeier
}

\authorfooter{
  \item
        Katherine J. Mimnaugh, Evan G. Center, Markku Suomalainen, Timo Ojala, and Steven M. LaValle are with the University of Oulu, Finland. \hfill \break E-mail: firstname.lastname@oulu.fi.
    \item
        Markku Suomalainen is with the VTT Technical Research Centre of Finland. E-mail: markku.suomalainen@vtt.fi.
    \item
        Israel Becerra, Eliezer Lozano, and Rafael Murrieta-Cid are with CIMAT, Mexico. Israel Becerra is also with Conahcyt, Mexico. E-mail: \{israelb, eliezer.lozano, murrieta\}@cimat.mx.
    \item
        Kara D. Federmeier is with the University of Illinois Urbana-Champaign, U.S.A. E-mail: kfederme@illinois.edu.
}

\abstract{
    In this paper, we show that Virtual Reality (VR) sickness is associated with a reduction in attention, which was detected with the P3b Event-Related Potential (ERP) component from electroencephalography (EEG) measurements collected in a dual-task paradigm. We hypothesized that sickness symptoms such as nausea, eyestrain, and fatigue would reduce the users' capacity to pay attention to tasks completed in a virtual environment, and that this reduction in attention would be dynamically reflected in a decrease of the P3b amplitude while VR sickness was experienced. In a user study, participants were taken on a tour through a museum in VR along paths with varying amounts of rotation, shown previously to cause different levels of VR sickness. While paying attention to the virtual museum (the primary task), participants were asked to silently count tones of a different frequency (the secondary task). Control measurements for comparison against the VR sickness conditions were taken when the users were not wearing the Head-Mounted Display (HMD) and while they were immersed in VR but not moving through the environment. This exploratory study shows, across multiple analyses, that the effect mean amplitude of the P3b collected during the task is associated with both sickness severity measured after the task with a questionnaire (SSQ) and with the number of counting errors on the secondary task. Thus, VR sickness may impair attention and task performance, and these changes in attention can be tracked with ERP measures as they happen, without asking participants to assess their sickness symptoms in the moment.

}


\teaser{
  \centering
  \includegraphics[width=\linewidth]{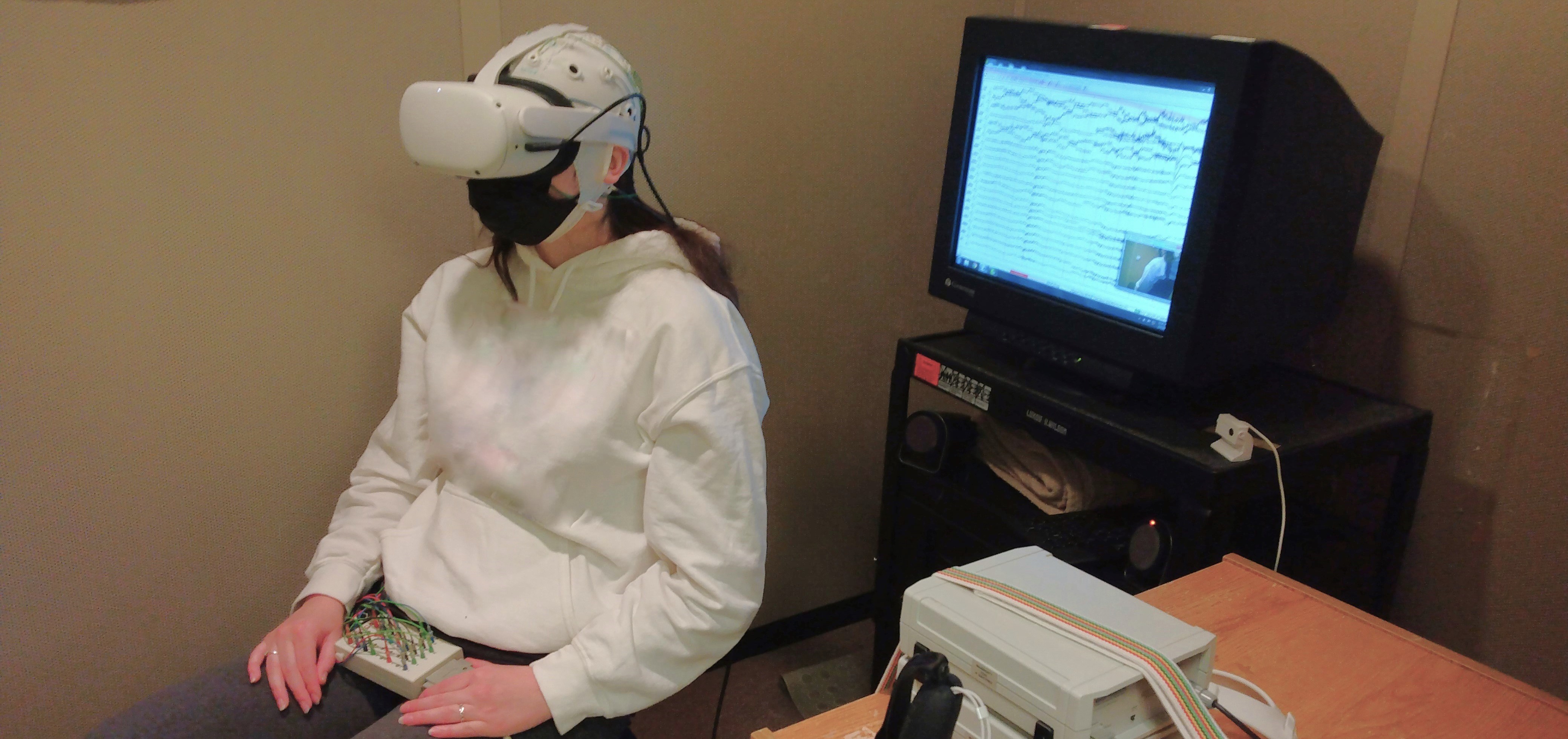}
  \vspace{-5mm}
  \caption{A participant is seated in the recording chamber wearing an EEG cap to measure her brain activity and an Oculus Quest 2 headset. A livestream of her brain activity is visible on the monitor behind her.}
  \label{fig:teaser}
}

\renewcommand{\manuscriptnotetxt}{}

\graphicspath{{figs/}{figures/}{pictures/}{images/}{./}} % where to search for the images

\usepackage{tabu}                      % only used for the table example
\usepackage{booktabs}                  % only used for the table example
\usepackage{lipsum}                    % used to generate placeholder text
\usepackage{mwe}                       % used to generate placeholder figures
\usepackage{textcomp}

\usepackage{mathptmx}                  % use matching math font

\begin{document}

%%%%%%%%%%%%%%%%%%%%%%%%%%%%%%%%%%%%%%%%%%%%%%%%%%%%%%%%%%%%%%%%
%%%%%%%%%%%%%%%%%%%%%% START OF THE PAPER %%%%%%%%%%%%%%%%%%%%%%
%%%%%%%%%%%%%%%%%%%%%%%%%%%%%%%%%%%%%%%%%%%%%%%%%%%%%%%%%%%%%%%%


\vspace{10mm}

% \firstsection{Introduction}

\maketitle

\section{Introduction} %for journal use above \firstsection{..} instead

Virtual Reality (VR) sickness is a set of symptoms similar to motion sickness that can result from the use of a VR Head-Mounted Display (HMD) \cite{Saredakis_Szpak_Birckhead_Keage_Rizzo_Loetscher_2020}. Not all people experience VR sickness, and users that are susceptible to discomfort do not always develop symptoms. However, when VR sickness does occur, it has been shown to detract from user experience \cite{somrak_humar_others_2019}, as well as cause symptoms that may persist after VR exposure has completed \cite{Duzmanska_Strojny_Strojny_2018, Szpak_Michalski_Saredakis_Chen_Loetscher_2019}. 
Consequently, as the access of HMDs for myriad uses in training, education, and therapeutics increases, interest in VR sickness has proliferated, with many articles published on the topic each year (see \cite{Chang_Kim_Yoo_2020, Saredakis_Szpak_Birckhead_Keage_Rizzo_Loetscher_2020} for review). 

However, there are still many open questions regarding the cognitive and behavioral impacts of VR sickness. Beyond simply causing unpleasant experiences that may discourage users from future use of the technology, VR sickness may interfere with users' ability to properly engage with content or perform well on tasks in VR. Previous work has found an increase in the error rate on tasks performed outside of VR after VR sickness was induced \cite{Wu_Zhou_Li_Kong_Xiao_2020}, and it is possible that, like motion sickness \cite{Matsangas_Mccauley_2013}, VR sickness could have detrimental effects on executive functioning and other cognitive resources. Therefore, there is reason to believe that VR sickness not only reduces the enjoyment of VR, but further seriously detracts from the ability of users to benefit from using VR for training, therapy, or education. Consequently, it is imperative to better understand how VR sickness impacts cognitive function, especially attention, which is critical for goal-directed behavior \cite{Lavie_Hirst_De_Fockert_Viding_2004}.

To address the important gap in knowledge regarding the impacts of VR sickness, we indexed attention via brain activity and behavior while participants experienced motion along paths in a virtual environment known to elicit varying levels of VR sickness (see Fig.~\ref{fig:teaser}). To our knowledge, this is the first study to utilize a dual-task paradigm during different VR sickness conditions to measure the impact of VR sickness on attention. Thus, this study provides an important contribution to the body of knowledge on VR sickness by introducing a novel method for measuring VR sickness effects on attention during VR HMD use and providing preliminary evidence that VR sickness has negative impacts on attention and task performance. Study materials and data are available online on OSF (osf.io/v9fst) and Harvard Dataverse (doi.org/10.7910/DVN/MENRFT).

%\vspace{2mm}

\section{Background}

%\vspace{1mm}

\subsection{Virtual Reality Sickness}

Like motion sickness, VR sickness is thought to be the result of sensory conflict, when incoming sensory stimulation does not match expectations for patterns of sensations that are based on previous experience \cite{Reason_Brand_1975}. Alternate theories, such as the multisensory reweighting hypothesis \cite{Oman_1990} and the postural instability theory \cite{Riccio_Stoffregen_1991} have also been proposed (see \cite{Stanney_Lawson_Rokers_Dennison_Fidopiastis_Stoffregen_Weech_Fulvio_2020} for review). In addition to an evolutionary explanation \cite{Treisman77}, sickness may be related to complications that arise from vestibular-autonomic reflexes \cite{Bogle_Benarroch_Sandroni_2022}, and it has been theorized that sickness is more likely to develop when there is greater stimulation to the semicircular canals compared to the otolith organs in the vestibular system \cite{Previc_2018}. VR sickness presents as a number of unpleasant sensations, such as nausea, fatigue, eyestrain, dizziness, and headache, as the result of using a VR HMD \cite{Rebenitsch_Owen_2016}. There are several factors that contribute in the development of symptoms. Elements related to the hardware of the HMD, like the weight of the headset, and individual characteristics of the users, like susceptibility to motion sickness, can play a role \cite{Chang_Kim_Yoo_2020}. Software is a significant contributor as well. Vection, the illusion of self-motion based on visual cues, can evoke Visually-Induced Motion Sickness (VIMS). Although the terms VR sickness, simulator sickness, and cybersickness can sometimes be used interchangeably, historically they referred to VIMS elicited with different displays. Here, we use the term VR sickness to denote symptoms resulting from the use of an HMD \cite{Lavalle_2023}.  In laboratory studies, VR sickness is often subjectively measured using self-report of the intensity of a number of physical symptoms \cite{Kennedy_Lane_Berbaum_Lilienthal_1993}. Objective measures of VR sickness include physiological arousal \cite{Dennison_Wisti_DZmura_2016}, gastrointestinal muscle activity \cite{Chang_Kim_Yoo_2020}, or brain activity \cite{Chang_Billinghurst_Yoo_2023}.

\subsection{Attention}

Attention is a multifaceted construct \cite{Hommel_Chapman_Cisek_Neyedli_Song_Welsh_2019}, which, broadly speaking, encompasses processes that serve to prioritize information relevant for accomplishing a goal \cite{Nobre_Kastner_Chapter_2014}. Attention is thus a property of perceptual and cognitive systems that provides a means of selecting from a set of possibilities \cite{Chun_Golomb_Turk-Browne_2011} or that serves as a set of weights and balances to prioritize how information is used  \cite{Narhi-Martinez_Dube_Golomb_2023}. One way of conceptualizing attention is as a finite pool of resources, or attentional capacity, that cognitive and perceptual processes collectively draw from \cite{Kahneman_1973}. When some processes use attentional resources from this this pool, what is left over is known as \textit{attentional reserve} \cite{Ghani_Signal_Niazi_Taylor_2020, Miller_Rietschel_McDonald_Hatfield_2011}. At low levels of effort among simultaneous processes, there may be sufficient capacity to accommodate all demands. If there are insufficient resources for all concurrent processes that require them, then attentional reserve may be depleted and performance may subsequently deteriorate \cite{Kahneman_1973, Jaquess_Gentili_Lo_Oh_Zhang_Rietschel_Miller_Tan_Hatfield_2017}. Although there are a wide variety of perceptual and cognitive processes that use attention across multiple modalities, space, and time, there is extensive evidence and consensus that prioritization is essential and that processing limitations occur \cite{Nobre_Kastner_Book_2014, Band_Jolicor_Akyrek_Memelink_2006}. Since processes utilizing attention are not directly observable, attentional changes can be quantified with a method called electroencephalography.

\subsection{Electroencephalography and Event-Related Potentials}

Electroencephalography (EEG) measures brain electrical activity in the form of post-synaptic potentials from populations of neurons that are recorded by electrodes placed on the scalp. Event-Related Potentials (ERPs) are the changes in voltage that occur in relation to a specific stimulus. In ERP experiments, researchers repeatedly expose subjects to a particular stimulus or type of stimulus while recording their brain activity. ERPs are then derived from the EEG signal by extracting segments that are time-locked to the stimuli of interest and compared across conditions of interest. Across decades of study, features of the ERP have been related to specific neural and cognitive functions. Different types of experimental models are designed to elicit particular ERP components and can be used to evaluate neural activity \cite{Luck_2014}. One of those experimental models, called the oddball paradigm, has been used for fifty years to study cognitive function \cite{Donchin_Ritter_McCallum_1978,Isreal_Chesney_Wickens_Donchin_1980, Mast_Watson_1968, Polich_2007}.

It has been well-established that an ERP component derived from the oddball paradigm, called the P3b, can be used as a measure of attentional reserve \cite{Ghani_Signal_Niazi_Taylor_2020}. The P3b has been associated with processing that is engaged when stimuli are evaluated with respect to a task goal, and, in particular, when attention is engaged to promote memory operations that revise active mental representations \cite{Polich_2007, Verleger_2020}. P3b amplitudes are sufficiently reliable to be used clinically for diagnosis of cognitive impairment \cite{Duncan_Barry_Connolly_Fischer_Michie_Naatanen_Polich_Reinvang_VanPetten_2009}, and the oddball paradigm can be presented in different sensory modalities. In an auditory oddball task, participants listen to a series of tones (standards) and are asked to keep track of the number of times they hear a less frequent tone that is, for example, a different pitch from the others (oddballs). Compared to the response to standards, the response to oddballs is characterized by a larger positivity observed at electrode locations over the center and back of the head, often peaking shortly after about 300 milliseconds (ms) post-stimulus onset. This response goes by several labels in the literature, but we will here refer to it as the P3b. A dual-task oddball paradigm involves asking participants to pay attention primarily to some other task while still keeping track of the number of oddballs as a secondary task. As mentioned in the previous section, it has been well established that tasks often "compete" for cognitive processing resources critical for task performance. Competition for these attentional resources creates a reciprocal relationship in a dual-task paradigm, such that performance on the primary task trades off with performance on the secondary task. Measures of the P3b can stand in for performance measures and provide a more direct, dynamic index of the availability of attentional resources, such that as attention is consumed by, for example, a more difficult primary task, P3b responses to secondary task oddballs show decreased amplitudes (for review, see \cite{Ghani_Signal_Niazi_Taylor_2020, Polich_2007, Kok_2001}).

\begin{figure*}[!t]
 \centering % avoid the use of \begin{center}...\end{center} and use \centering instead (more compact)
 \vspace{-2mm}
\includegraphics[width=\textwidth]{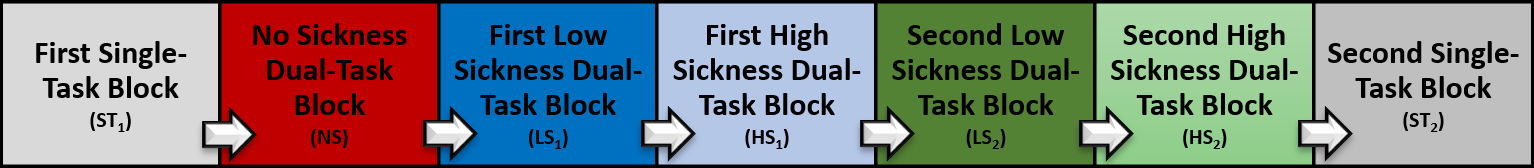}
\vspace{-5mm}
 \caption{The order of stimuli during the study. Each auditory oddball paradigm block consisted of 120 tones (each called a trial) with a varying number of infrequent higher pitch oddball tones interspersed between recurring lower pitch standard tones.}
 \label{fig:Methods}
 \vspace{-4mm}
\end{figure*}

\subsection{EEG and ERPs in Virtual Reality}

Despite the risk of the VR HMD disrupting the EEG recordings either via electrical or physical interference between the two devices, previous research utilizing a combination of EEG and VR has found suitability in their simultaneous deployment. Aksoy et al. \cite{Aksoy_Ufodiama_Bateson_Martin_Asghar_2021} compared sensory components (N1 and P1 responses) and the P3b within subjects completing an n-back working memory task on both a computer screen as well as in VR. They found similar timing of ERP component waveforms across the desktop and HMD conditions, and similarity in the P3b in mean and peak amplitudes, and for peak latency, across the two screen conditions. Harjunen and colleagues \cite{Harjunen_Ahmed_Jacucci_Ravaja_Spape_2017} compared early visual and sensory processing components for a bimodal lateralized oddball task in embodied, not embodied, and control conditions inside and outside a VR headset. They found that for the visual evoked potentials during the P3b, there was less noise and more signal in the conditions with the VR headset on as compared to the control condition with the headset off. They also found that fewer trials needed to be rejected for artifacts when the subject was wearing the HMD, and that the HMD had no systematic influence on any of the ERPs that were collected.

The assessment of attention with virtual reality has also been established in the literature. In the area of therapeutics, several studies have explored VR-adapted neuropsychological assessments for diagnosis and VR-based therapies for treatment of Attention Deficit Hyperactivity Disorder (ADHD) \cite{Bashiri_Ghazisaeedi_Shahmoradi_2017}. Other beneficial impacts for VR use on attention were investigated by Chung and colleagues \cite{Chung_Lee_Park_2018}, who explored the restorative effects of nature exposure in virtual reality on attention using a dual-task auditory oddball paradigm. Subjects were instructed to ignore the auditory stimulus while viewing panoramic videos of natural scenes or fireworks in VR, and ERP responses related to preattentive pattern detection (MMN) and sensitivity to attentional capture by novelty (P3a) were evaluated. The researchers found a reduction in the amplitude of the P3a, which reflects a decrease in attentional capture from the distracting stimuli, in the group of participants that experienced greater restoration from the nature exposure.

Although it has been shown that motion sickness not induced by VR can have deleterious impacts on cognitive function \cite{Matsangas_Mccauley_2013}, there are still unanswered questions regarding whether these same effects would be found with VR sickness, and whether VR sickness would affect attention specifically. There is reason to believe that motion through VR environments would have an impact on attention as motion through environments in the real world has shown attentional effects \cite{Ladouce_Donaldson_Dudchenko_Ietswaart_2019}, and VR sickness has been found to display similar timing and symptoms as motion sickness \cite{Gavgani_Walker_Hodgson_Nalivaiko_2018}. Additionally, vection has been found to be associated with changes in attention. Stróżak and colleagues \cite{Strozak_etal_2016} used a dual-task visual oddball paradigm wherein subjects viewed a pattern of X's and O's against a background with black and white vertical strips, which were either static or moved horizontally to induce vection. They found a reduction in P3b amplitudes during the conditions in which there was a vection-inducing background as compared to the stationary background, and an increase in P3b peak latency when vection was induced across the entire visual field. They also found that subjects had slower reaction times when the targets occurred after vection was induced. However, they did not measure whether subjects experienced any sickness symptoms. Since the experience of vection does not always result in the development of VR sickness \cite{Keshavarz_Riecke_Hettinger_Campos_2015}, the question remains whether VR sickness from immersion in a VR HMD has an impact on attention.

Preliminary evidence of the potential impact of VR sickness on attention can be found in a study by Wu and colleagues \cite{Wu_Zhou_Li_Kong_Xiao_2020}. Twenty male participants completed a visual oddball task before and after completing a forty-minute VR sickness-inducing navigational task through a space station in an HTC Vive HMD. The researchers found a significant reduction in P3b amplitude and an increase in P3b latency after the VR condition as compared to before it. They also found a significant increase in the number of errors in the post-test as compared to the pre-test. Although their study did not evaluate the P3b during the experience of VR sickness, they argue that symptoms have been shown to persist after the VR stimulus has ended, and thus that the changes in the P3b that they found were reflective of the influence of the VR sickness. Nevertheless, they argue for the need to investigate VR sickness on cognitive function during VR exposure.

\section{Methods}

\subsection{Study Overview and Aims}

This exploratory EEG study investigated how increasing levels of VR sickness impacted participants' attention as they viewed a VR environment. To validate our methods against previous work, we first measured subjects' attention outside the VR headset with a single-task oddball paradigm (see Fig.~\ref{fig:Methods} for an overview of the study progression). Then, subjects put on the VR headset and viewed a static gallery at the start of a virtual museum, which provided an interesting immersive environment with artwork and statues. The stationary VR museum gallery allowed for ERP measures in a very basic dual-task oddball paradigm (the no sickness condition). To create predictable levels of VR sickness symptoms from multiple exposures to the same environment in VR, we selected stimuli that had been previously shown to result in different intensities of VR sickness. In our previous work \cite{Becerra_Suomalainen_Lozano_Mimnaugh_Murrieta-Cid_Lavalle_2020}, we found piecewise linear paths through virtual environments less sickening than smooth curvilinear paths. For this study, we used the same environment, except the low sickness condition path was half-speed so that the two sickness condition museum tours were approximately similar in length, and the paintings were more evenly distributed throughout the museum. The museum tours can be seen in the video and the Unity project downloaded from OSF (osf.io/v9fst). The museum tours were designed to potentially create mild (low sickness condition) and moderate (high sickness condition) sensations of VR sickness based on the path trajectories used while traversing the environment. In the low sickness condition, the path moved in straight lines down the center of the hallways in which the artwork was displayed, making only two 90-degree turns along the route. In the high sickness condition, the path took a circuitous route, moving closer to the walls and paintings in some areas of the museum, and making many large and small degree turns, as rotational motion in particular has been shown to increase VR sickness \cite{Chang_Kim_Yoo_2020}. We anticipated that rising levels of unpleasant internal bodily sensations would involuntarily capture increasing amounts of attention, thus depleting attentional resources available for other tasks. Thus, we hypothesized that increasing levels of VR sickness, as measured by increases in scores on the Simulator Sickness Questionnaire, would be associated with decreases in attentional reserve, as measured by smaller effect (oddball minus standard) mean amplitudes of the P3b.

\subsection{Participants}

Data were collected at the University of Illinois on participants that included members of the university community and the surrounding area. The study was designed in line with the principles of the Declaration of Helsinki and approval from the local Institutional Review Board (\#22519) was obtained. An additional COVID-19 Safety Plan protocol was approved, and the study was conducted in-person on-campus in accordance with those guidelines. Subjects were recruited through flyers posted around campus and invitational emails to individuals who had previously participated in research studies at the university. Recruitment materials described the study as being designed to elicit mild to moderate sensations similar to motion sickness, and individuals that were pregnant or very sensitive to motion sickness were encouraged not to sign up for the study. To participate in the study, participants were required to have normal or corrected-to-normal vision and hearing, no colorblindness, no previous diagnosis of epilepsy or other seizure disorder, and no history of skull fractures. Participants were also asked not to use drugs or alcohol in the 24 hours prior to participation. 

Twenty-nine people (17 women) ages 18 to 50 (1 unreported, $M = 23.11,~SD = 6.87)$ participated in the research study. Forty-eight percent (n = 14) of the sample had never used  a VR HMD previously, 35\% (n = 10) had only used a VR HMD once or just a couple of times ever, 10\% (n = 3) used a VR HMD once or twice a year, and 7\% (n = 2) used a VR HMD once or twice a month. Regarding the frequency with which they played computer games, 7\% never played, 7\% played once or just a couple of times ever, 35\% played once or twice a year, 17\% played once or twice a month, 10\% played once or twice a week, 14\% played several times a week, and 10\% played every day.

\subsection{Measures} \label{Measures}

\subsubsection{Attention Measurement} 

In this experiment, we measured changes in attentional reserve with a dual-task auditory oddball paradigm by comparing the difference in P3b effect mean amplitudes (oddball minus standard) across blocks and conditions. The auditory oddball paradigm consisted of more frequent lower tone (440 Hz) standards interspersed with less frequent higher tone (660 Hz) oddballs presented in a semi-random order with a minimum of at least one standard tone separating each of the oddballs. The percentage of standard tones varied from 71 to 77\% depending on the block. The tones lasted for 200 ms with a randomly generated inter-stimulus interval (ISI) between 600 and 1000 ms. Each time a tone was presented counted as one trial. A total of 120 trials were presented in each block (group of trials), but the first six tones (pre-task standards) were coded separately and did not count towards the 114 trials in each block used in ERP analysis. These first six rejected trials allowed subjects a little bit of time to prepare for mentally counting the oddballs after the tones began to play. The oddball task was completed seven times for a total of seven blocks (see Fig.~\ref{fig:Methods}). The order of blocks was fixed to increase power given the limited sample size and because the accumulation of VR sickness meant that counterbalancing could have washed out effects for any low sickness block collected after a high sickness block. To enhance power to detect a sickness condition effect, we repeated the low and high sickness blocks, and combined them to create the low and high sickness conditions. The stimuli were presented using the software program Presentation and a modified version of the ERP CORE package available online from Kappenman et al. \cite{Kappenman_Farrens_Zhang_Stewart_Luck_2021}. 

During the single-task blocks, subjects were instructed to sit still and look at a fixation cross while silently counting the number of oddball tones. During the dual-task no sickness block, participants were instructed to look around at the museum by turning their chair up to 90 degrees in either direction while silently counting the tones. During the dual-task sickness blocks, subjects were instructed that their primary task was to view and enjoy the artwork and architecture in the virtual museum while they silently counted the number of oddball tones that occurred during the tour. They were told that if it became too difficult to look around at the museum and count the tones, then they should prioritize viewing the museum (about which they would be tested later) even if they missed counting some tones. At the end of each of the approximately two minute blocks (each block was a slightly different length due to variable ISIs), participants were asked to report the number of oddballs that they heard. The number of oddballs in each block ranged from 27 to 34, with the number of oddballs per block type held consistent across participants. To motivate participants to primarily attend to the virtual tours, they were given a quiz at the end of the study regarding the artwork they had seen. Subjects were presented with pictures of sixteen paintings and sculptures, of which nine were present in the museum and seven were new pieces of a similar style. Participants were also asked open-ended questions regarding their allocation of attention and if their attention changed during the study. Their answers to those questions were not analyzed here, but are available on OSF (osf.io/v9fst).

\subsubsection{Sickness Measurement} \label{SSQ}

The Kennedy et al. \cite{Kennedy_Lane_Berbaum_Lilienthal_1993} Simulator Sickness Questionnaire (SSQ) was used to measure participants' VR sickness symptoms throughout the study. The questionnaire consists of 16 symptoms about which users rate the severity of their experience as either none, mild, moderate, or severe. The questionnaire asks how much the person is experiencing that symptom right now, but after the VR stimuli participants were instructed to rate their symptoms as they were experiencing them now or if they had experienced them at any point during the museum tour that they had just completed. The first SSQ was administered at the beginning of the study before the EEG cap setup, as a baseline measurement of any symptoms (like fatigue or headache) that were already present before the VR sickness induction. The subsequent four SSQs were administered immediately after each of the low and high sickness blocks. The questionnaires were scored by computing the weighted sum of the symptoms for each of the Nausea, Oculomotor, and Disorientation subscales, as well as for the Total Score. Subjects were also asked if they had closed their eyes at any point during the experiment, or if the lenses became foggy at any point to screen for potential issues with correctly completing the tasks.

\subsection{Equipment}

The virtual museum tours were created in Unity 3D and displayed with the Oculus Quest 2, which has a single fast-switch LCD screen with a resolution of 1832 by 1920 pixels per eye and a 72 Hz refresh rate. The Unity executable was run in real-time and launched through an Oculus Link cord from a desktop computer. A circular flexible thermoplastic polyurethane Quest 2 head back pad was fitted underneath the Oculus straps to reduce direct pressure on the occipito-parietal electrodes, which could have caused interference in recording the brain activity. Altec Lansing BXR1220 desktop speakers were located behind the participant and played the audio tones for the oddball task. 

The EEG data were collected with a Brain Products BrainAmp DC amplifier with a 0.016 to 250 Hz bandpass filter and sampled at 1000 Hz. Sixteen passive Ag/AgCl electrodes, a subset from an equidistantly spaced layout (seen in the inset of Fig. \ref{fig:Difference_Waves}), were used. One electrooculogram (EOG) sensor was placed on the infraorbital ridge below the left eye and two EOG electrodes were placed on the outer canthus of the left and right eyes to assist with artifact rejection. The electrodes were referenced online to an electrode placed on the left mastoid, and re-referenced offline to the average of the electrodes on the left and right mastoids. Impedances were kept under 5 k$\Omega$.

\subsection{Procedure}

When participants first arrived for the study, their eligibility to participate in the study was confirmed. Then, they read information about the study and signed a consent form to participate. They were verbally briefed on the nature of the tasks and their rights as research participants were explained to them. Next, their interpupillary distance (IPD) was measured by having the subjects focus on a point in the distance and using a tape measure to determine the distance between the center of their pupils. Their IPD was used to adjust the VR lenses, and they tried on the HMD to ensure fit and clarity. They also practiced navigating through some menus to familiarize themselves with the HMD. Participants then completed a baseline sickness measurement. Participants were fitted with the EEG cap and then moved to a swivel chair in an electrically shielded chamber and impedances for each electrode were checked. The chair height was adjusted and they were instructed to use their feet to turn the chair around as opposed to using their neck muscles to move their head, which would create noise in the reference electrodes placed on the left and right mastoids.

The EEG recording began with the first single-task oddball paradigm. A fixation cross was placed on the wall at a comfortable height for the user. Participants were instructed to sit still while staring at the fixation cross, and to silently count the number of higher pitch tones that occurred less frequently, which were then reported to the experimenter at the end of the block. Then, the VR HMD was placed on their head and the virtual museum was launched. Subjects viewed a gallery at the start of the museum with paintings on the walls around them which they could turn to look at but not move towards. For the dual-task no sickness condition block, subjects were instructed to look around at the museum by rotating their chair while silently counting the number of oddball tones. At the end of the block, they again verbally reported the number of oddballs to the experimenter. 

Subsequently, the VR sickness condition blocks began. Participants were instructed that their primary task was to view and enjoy the art and architecture in the museum, about which they would later be tested, but that they should still try to be as accurate as possible in counting the number of oddballs. They were told they would go through the museum four times so they would have more than one chance to view all of the art, and that they would be given a quiz at the end of the experiment about the artwork present in the museum. The four dual-task oddball paradigms were presented in fixed order: low sickness, high sickness, low sickness, and high sickness. During each block, the tour was initiated by the experimenter and then the oddball tones began to play a couple of seconds later. Immediately after each tour ended, the subjects were asked to report the number of oddballs that they heard, and then they moved the HMD off of their face slightly to answer the SSQ questionnaire on a separate laptop. The HMD was put back into place and the electrodes were checked for proper connection before each of the following museum tours began. After the four sickness condition blocks and SSQs were completed, the HMD was removed and participants completed one final single-task oddball block. They were then moved out of the chamber and the electrodes were removed while they completed the post-experiment questionnaire and artwork quiz. The study took about two hours to complete. After they had finished, they were debriefed on the nature of the study, and given compensation of 12 USD per hour in gift cards for their participation.

\subsection{EEG Preprocessing}

\subsubsection{Independent Component Analysis}
Eye movements and blinks create electrical deviations that are separate from those elicited by the brain. Independent Component Analysis (ICA) is a commonly used procedure to separate these artifact signals from the brain activity \cite{Hyvarinen_Oja_2000, Jung_Makeig_Humphries_Lee_Mckeown_Iragui_Sejnowski_2000}. However, given that ICA has primarily been used on datasets collected when head and eye movements are restricted, there was some uncertainty as to the best method for dealing with artifacts in the context of VR. Therefore, the data were preprocessed using three different methods that varied based on the filter settings, the ICA algorithms, and whether or not the data were epoched before ICA decomposition. After comparison of ERP waveforms and statistical analysis of the mean amplitudes across the time window of interest, the results showed similar patterns. Here we report the preprocessing pipeline that was selected based on the recommendations from an article on ICA for free viewing tasks \cite{Dimigen_2020} and from which some code made available online for ICA preprocessing was used \footnote{Code from Olaf Dimigen: \url{https://github.com/olafdimigen/OPTICAT/blob/master/opticat_script.m}}. The Matlab preprocessing script for EEGLAB and results from the three pipelines are available on OSF (osf.io/v9fst).

\subsubsection{Preprocessing Pipeline}
EEG data were preprocessed in Matlab (Mathworks) using the EEGLAB software toolbox \cite{Delorme_Makeig_2004} following the guidelines from Luck \cite{Luck_2014}, and downsampled to 250 Hz for analysis. Prior to ICA, the data were high-pass filtered following Dimigen \cite{Dimigen_2020} with a zero-phase Hamming-windowed sinc FIR filter at 2 Hz (the passband edge; note that the -6 dB half-amplitude cutoff is 1 Hz). The data were divided into one second epochs from 200 ms before stimulus onset to 800 ms after stimulus onset, and mean-center corrected \cite{Groppe_Makeig_Kutas_2009}. The extended Infomax ICA algorithm in EEGLAB was used to calculate the ICA weights and ICA sphering matrix. These values were then transferred back to the downsampled and unfiltered raw EEG data. The 16 scalp channels, three EOG channels, and one mastoid channel were used to compute 20 independent components for each subject. One horizontal (the difference between the left and right outer canthus eye channels) and one vertical (the difference between the left prefrontal scalp channel and the lower left eye EOG channel) bipolar eye channels were created to assist with the detection of eye-related artifacts in the data. The data were then bandpass filtered with a second order IIR Butterworth filter with a 0.2 to 30 half-amplitude cutoff and 12 dB/oct roll-off \cite{Widmann_Schroger_Maess_2015}. ICLabel was used to classify the ICs by the compositional vectors of types of artifacts (for example, percentage of the scalp data variance accounted for which is classified as eye activity, muscle activity, and brain data) \cite{Pion-Tonachini_Kreutz-Delgado_Makeig_2019}. From the resulting list of ICs, components with high IC class probabilities as eye activity or which displayed scalp topographies characteristic of eye movement and blink-related artifacts were inspected by eye and compared against the raw EEG signal and bipolar channels for each subject to confirm which ICs included eye-related signal interference. As few ICs as possible were selected to remove ocular artifacts while minimizing any potential excision of brain activity. This resulted in two artifact ICs removed for 24 subjects, three artifact ICs removed for three subjects and four artifact ICs removed for two subjects. The remaining 16 to 18 ICs were then combined back into the data matrix.

\subsubsection{ERPs}

After ICA correction, the epoched EEG data were baseline corrected by subtracting the 200 ms pre-stimulus activity from the 800 ms post-stimulus epoch. To reject any additional periods contaminated by artifacts after ICA removal of eye-related activity, a simple voltage threshold function was used to reject any epochs for all of the electrodes in the ROI that had a voltage less than -100 \textmu V or greater than 100 \textmu V during the one second time window around stimulus onset. EEG data from 20 of the participants had no additional epochs rejected, and the data from the remaining nine participants had between 1 and 12 trials rejected (0.1\% to 1.5\% of the total number of 798 trials, respectively). 

The EEG data were then averaged across blocks for each subject using the ERPLAB toolbox \cite{Lopez-Calderon_Luck_2014}. Measurements of the mean amplitude for each stimulus type (oddball or standard) in each of the conditions (single-task, dual-task no sickness, dual-task low sickness, and dual-task high sickness) were calculated for the time window between 250 ms and 450 ms after stimulus onset. This time window was selected based on extensive previous research on P3b effects (see \cite{Verleger_2020} for review) as well as visual inspection of the average waveform combining participants and conditions. The mean amplitude measurements were averaged across eight central-posterior electrodes where P3b amplitude effects are typically most prominent \cite{Polich_2007}, as seen in the inset to Fig.~\ref{fig:Difference_Waves}, to create an analysis Region-Of-Interest (ROI) similar to Hubbard and Federmeier \cite{Hubbard_Federmeier_2021}. 

\section{Results}

Data analysis was conducted using SPSS Statistics (IBM) and R software. Exploratory analyses were conducted with two-sided tests, alpha level set at .05, and Confidence Intervals (CIs) of 95\%. All data were tested for normality using Shapiro-Wilk tests, and non-Gaussian data were analyzed with non-parametric tests. Analyses were run with data from all 29 participants. EEG and SSQ data from the study are available on OSF and Harvard Dataverse (doi.org/10.7910/DVN/MENRFT).

To provide motivation for subjects to attend to the primary task of paying attention to the artwork and statues in the virtual museum, subjects were given a quiz at the end of the study asking if the images of artwork and statues presented to them had been present in the virtual museum. Overall, subjects performed very well on the quiz, with an average of less than one error per subject ($M = 0.966,~SD = 1.017$). The artwork quiz scores were not included in analyses.

\subsection{VR Sickness}

\begin{figure}[t]
 \centering % avoid the use of \begin{center}...\end{center} and use \centering instead (more compact) 
 \includegraphics[width=\columnwidth]{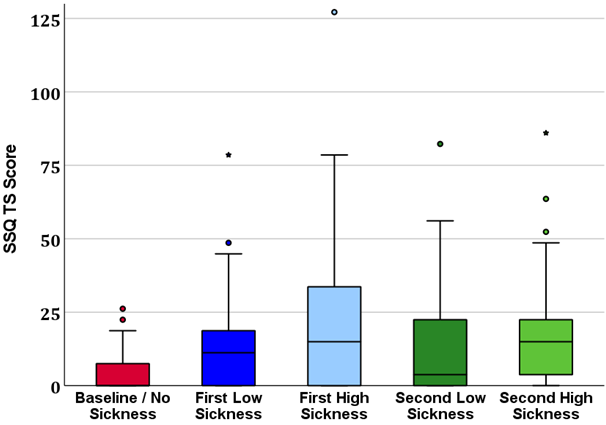}
 \vspace{-5mm}
 \caption{SSQ weighted Total Scores by block. Colors correspond with the waveforms generated during the same blocks below in Fig.~\ref{fig:Difference_Waves_AllBlocks}.}
 \label{fig:SSQ_box}
 \vspace{-4mm}
\end{figure}

The SSQ questionnaire was administered five times: a baseline / no sickness (NS) at the beginning of the study, after the first low sickness block (LS$_1$), after the first high sickness block (HS$_1$), after the second low sickness block (LS$_2$), and after the second high sickness block (HS$_2$). All scores were calculated as the weighted sum of the intensity of symptoms \cite{Kennedy_Lane_Berbaum_Lilienthal_1993}. Boxplots of the SSQ scores are presented in Fig.~\ref{fig:SSQ_box}.

\subsubsection{SSQ Score Differences by Block}
As SSQ scores are not normally distributed, the difference in SSQ weighted Total Score (TS) between the five sickness conditions was compared using a related-samples Friedman's two-way ANOVA. There was a statistically significant difference in SSQ TS across blocks, $\chi^2(4, \textit{N} = 29) = 24.193,~p < .001,~W = 0.209$. Post-hoc pairwise comparisons revealed no significant difference between the baseline / no sickness and low sickness blocks or between the sickness blocks compared with each other ($p > .05$). However, both high sickness condition SSQ TS were significantly greater than the baseline / no sickness scores, HS$_1 (p < .001)$ and HS$_2 (p = .001)$.

\subsubsection{SSQ Score Differences by Condition}
To examine sickness by condition, SSQ TS for both low sickness blocks (LS$_1$ and LS$_2$) %, $M = 13.864,~SD = 18.455$) 
and high sickness blocks (HS$_1$ and HS$_2$) %, $M = 20.957,~SD = 23.929$) 
were averaged and then compared against the baseline (NS) %, $M = 5.030,~SD = 7.498$) 
with a Friedman test. Again, there was a statistically significant difference between conditions, $\chi^2(2, \textit{N} = 29) = 22.828,~p < .001,~W = .394$, with post-hoc pairwise comparisons showing high sickness significantly greater than no sickness, $p < .001$, low sickness significantly greater than no sickness, $p = .030$, and high sickness significantly greater than low sickness, $p = .049$.

\begin{figure}[!b]
 \centering % avoid the use of \begin{center}...\end{center} and use \centering instead (more compact)
 \includegraphics[width=\columnwidth]{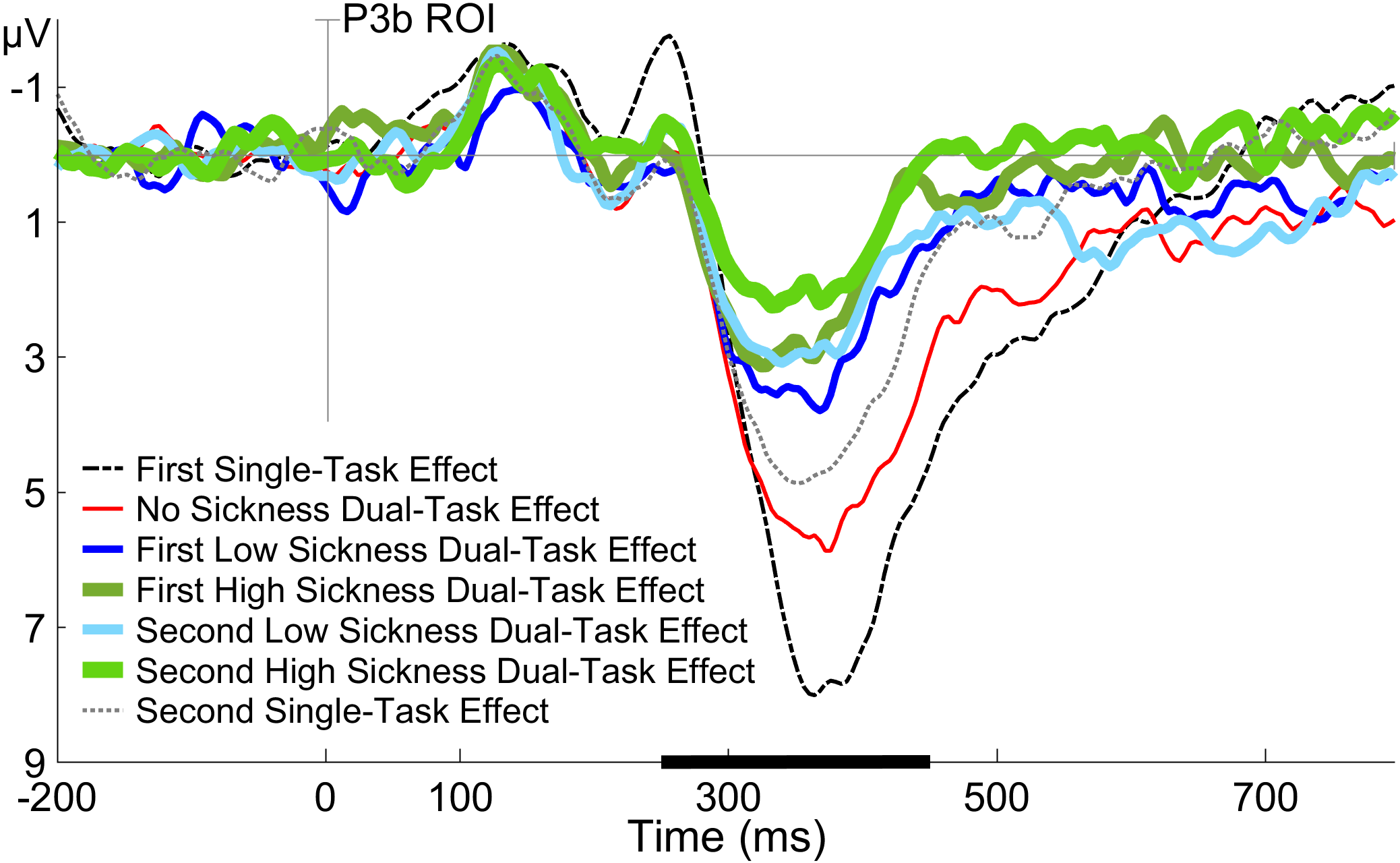}
 \vspace{-5mm}
 \caption{ROI P3b difference effects by blocks. Waveforms are differentiated by color, style, and line thickness. The thick black line on the x-axis denotes the time window for analyses, from 250 ms to 450 ms.}
 \label{fig:Difference_Waves_AllBlocks}
\end{figure}

\subsection{ROI P3b Effect Mean Amplitudes} \label{P3b_Results}

Eight electrodes in the ROI were averaged together to create a single ROI channel from which the mean amplitude measurements were taken. The P3b effect was calculated as the subtraction (oddballs minus standards) of the mean amplitudes in \textmu V from 250 to 450 ms post-stimulus.

\subsubsection{No VR Condition versus No Sickness VR Condition}

To test if using an HMD would have an impact on the P3b, we compared the P3b effect mean amplitude of the ROI between the first single-task (outside of VR) condition $(M = 4.53,~SD = 2.67)$ with the no sickness dual-task (in VR but no motion) condition $(M = 3.82,~SD = 3.04)$ that immediately followed. As the effect mean amplitudes were normally distributed, a paired-samples t-test was used. There was not a statistically significant difference between the means of the two conditions, $t(28) = 1.588, p = .124, d = 0.295, 95\%$ CI = [-0.205, 1.616], and thus the use of VR by itself did not result in a notable reduction in attention.

\subsubsection{First Single-Task versus Second Single-Task}

Since there was a possibility that the sickness symptoms induced during the VR stimuli could persist and continue to have an effect on attention after the VR tasks were completed, we examined the change between the first single-task block at the beginning of the experiment $(M = 4.53,~SD = 2.67)$ and the second single-task block after all the VR sickness conditions were completed $(M = 3.11,~SD = 2.44)$. A paired-samples t-test showed that the ROI P3b effect mean amplitudes were significantly reduced after the VR stimuli as compared to before them, $t(28) = 2.623, p = .014, d = .487, 95\%$ CI = [0.310, 2.521]. Thus, although the single-task paradigm was the same before and after the VR stimuli, there was a reduction in attention that may have been due to VR sickness symptoms that had not yet abated.

However, to further test whether the change in single-task P3b effect mean amplitudes was related to accumulated sickness, the change in SSQ TS scores and P3b effect mean amplitudes from the beginning to the end of the study (HS$_2$ - NS) were compared. The correlation was not significant, $r = .212, p = .271$, and thus it is possible that some other factor like boredom completing the task for a seventh time led to this decrease in amplitudes.

\begin{table}[!t]
 \centering % avoid the use of \begin{center}...\end{center} and use \centering instead (more compact)
 \includegraphics[width=\columnwidth]{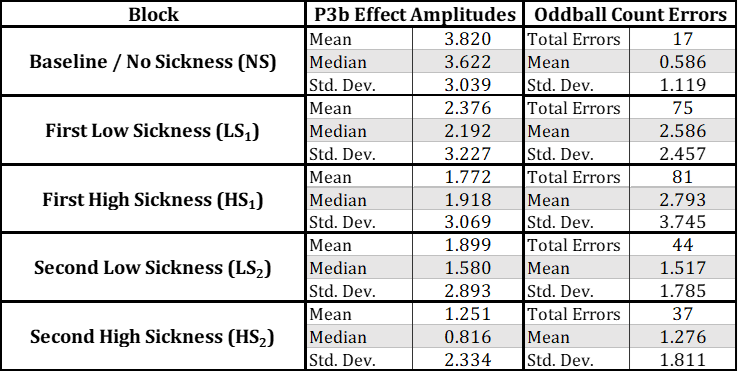}
 \vspace{-5mm}
 \caption{P3b Effect Mean Amplitudes and Oddball Count Errors by Block. 
}
 \label{fig:SSQ_scores}
 \vspace{-5mm}
\end{table}

\begin{figure}[!b]
 \centering % avoid the use of \begin{center}...\end{center} and use \centering instead (more compact)
 \includegraphics[width=\columnwidth]{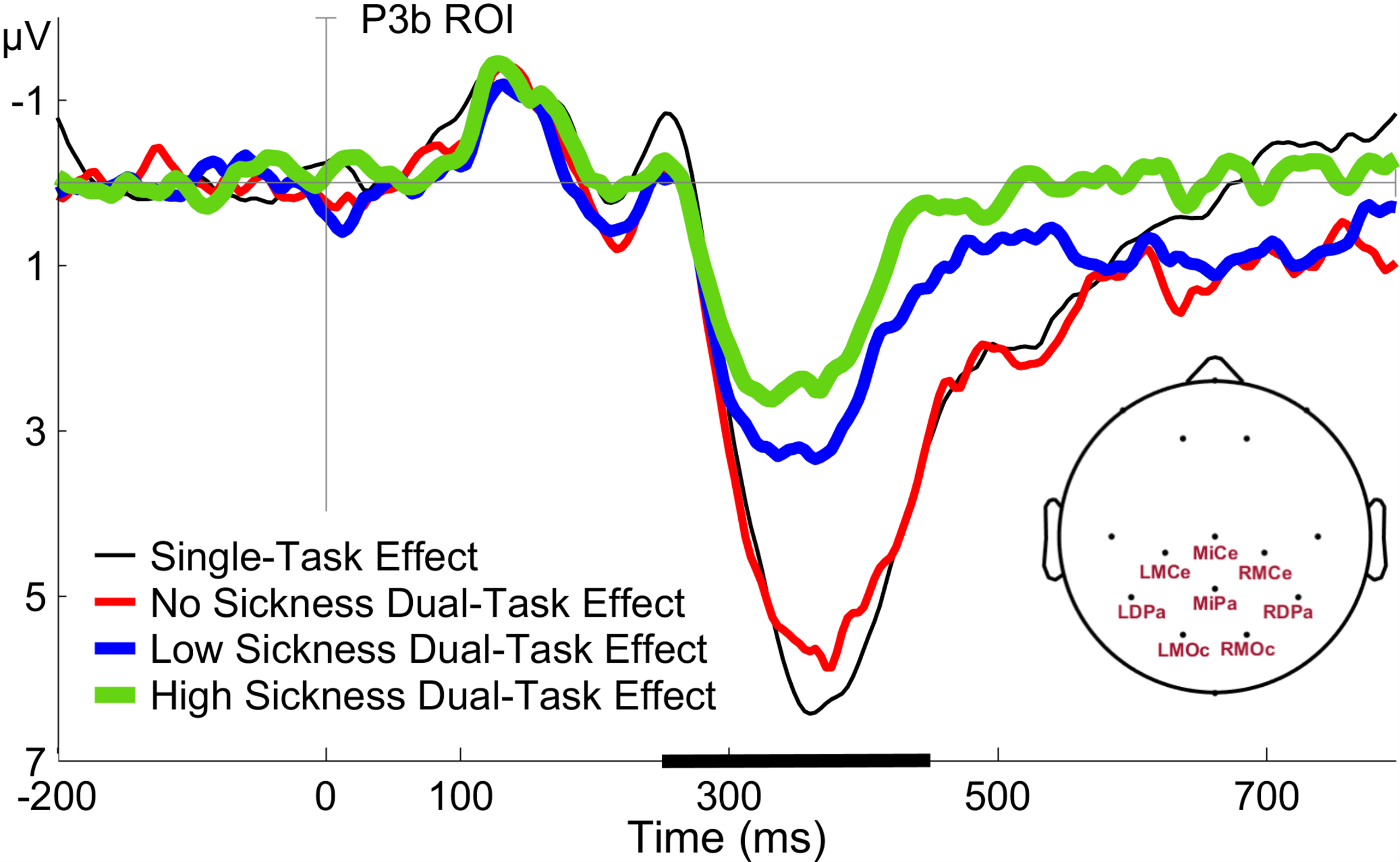}
 \vspace{-6mm}
 \caption{ROI P3b difference effects by condition. The inset shows a top-down view of a head with the central-posterior P3b Region-Of-Interest (ROI) electrodes from which these waveforms are derived in red.}
 \label{fig:Difference_Waves}
\end{figure}

\subsubsection{Effect Mean Amplitude Differences by Block} \label{P3b_Effects_Block}

To investigate the differences in attention depending on the level of sickness induced, a one-way repeated measures ANOVA was used to compare ROI P3b effect mean amplitudes across the five sickness dual-task blocks (waveforms shown in Fig.~\ref{fig:Difference_Waves_AllBlocks}; descriptives per block in Table~\ref{fig:SSQ_scores}). There was a significant effect of block, $F(4, 112) = 6.533, p < .001, \eta^2 = .189$. Post-hoc pairwise comparisons showed a statistically significant reduction in attention between the no sickness block and all of the sickness blocks, with the first low sickness block ($p = .001$), the first high sickness block ($p < .001$), the second low sickness block ($p = .001$), and the second high sickness block ($p < .001$) displaying attenuated P3b effect mean amplitudes as compared to the no sickness block. There were no significant differences between any of the low and high sickness blocks when compared with each other, except between the second high sickness block and the first low sickness block, which showed a significant decrease in P3b amplitude  $(p = .046)$.

\begin{figure}[!t]
 \centering % avoid the use of \begin{center}...\end{center} and use \centering instead (more compact)
 \includegraphics[width=0.9\columnwidth]{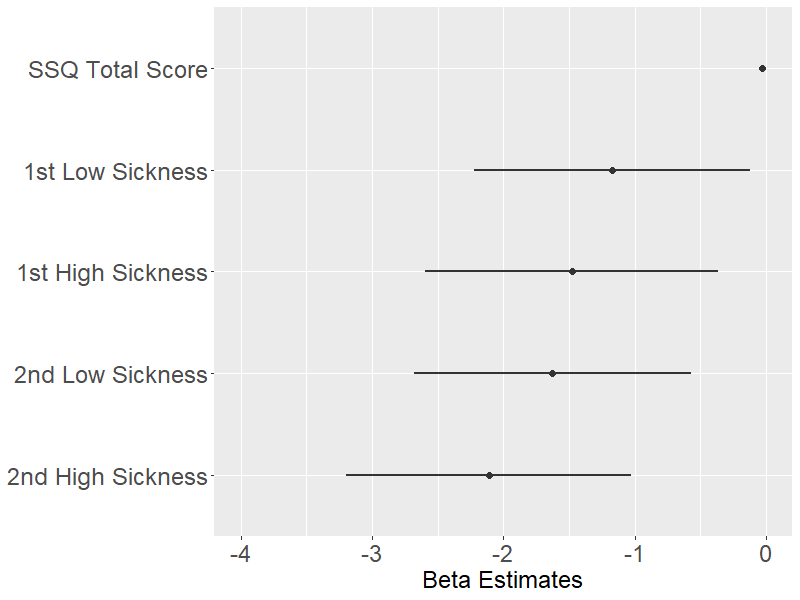}
 \vspace{-2mm}
 \caption{Beta point estimates (regression coefficients) for full model predictors with 95\% CIs.}
 \label{fig:Beta_points}
 \vspace{-4mm}
\end{figure}

\subsubsection{Effect Mean Amplitude Differences by Condition} \label{P3b_Effects_Condition}

As the P3b effect mean amplitudes were not normally distributed, a related-samples Friedman's two-way ANOVA was used to examine the differences in attentional reserve between sickness levels across the dual-task conditions (see Fig.~\ref{fig:Difference_Waves}) of no sickness ($Mdn = 3.622$), the average of both low sickness blocks ($Mdn = 1.673$), and the average of both high sickness blocks ($Mdn = 1.154$). The difference between sickness conditions was significant, $\chi^2(2, \textit{N} = 29) = 16.345,~p < .001,~W = .282$. Post-hoc comparisons showed a decrease in the P3b effect mean amplitudes from no sickness to low sickness $(p = .006)$ and from no sickness to high sickness $(p < .001)$, but not from low sickness to high sickness $(p = .237)$.

\subsection{Modeling Relationships} \label{HLM_section}

We were interested in investigating the relationships among task performance (the number of errors in reporting how many oddball beeps were heard in each block), sickness (SSQ scores), the measure of attentional reserve (ROI P3b effect mean amplitudes), and demographic factors (gender, VR use, computer gaming use). However, most of these measures were acquired from the same individual after every block, making correlations inappropriate as they cannot account for the dependence between the data points deriving from the same individuals. We instead chose a multi-level modeling (MLM) approach to test whether differences in task performance, sickness, or demographics would be associated with the changes in attention observed in the study. 

\subsubsection{Attentional Allocation}

We reasoned that while we intended for certain blocks to be more sickening, individual variability in sickness might lead to effects being masked at the block level. Upon examination, indeed there were some participants who scarcely ever experienced sickness, while some others were continuously sick throughout the duration of the experiment. To properly address this variability, we fit the data using hierarchical linear modeling (HLM) via the R statistical software package `lme4' \cite{Bates_et_al_2015} to test whether the degree of sickness in each condition predicted P3b effect ROI mean amplitudes, both on its own, and over and above the effect contributed by the path trajectories of each condition when both sets of predictors were included. To do so, we constructed and compared five nested models with the P3b effect mean amplitudes as our outcome variable and a random intercept for participants to account for multiple responses deriving from the same participants in the repeated measures design. These models included a "null" model with only a fixed effect intercept (which can be interpreted as a "baseline" case representing the null hypothesis that none of our data predict the allocation of attentional resources), a "sickness only" model which added a fixed effect slope for a sickness severity predictor (on a per block per participant basis, as indicated by SSQ TS), a "condition only" model which contained only fixed effect slopes for block types as predictors, a "full" model which added fixed effect slopes for both sickness severity predictors and block types, and a "full + demographics" model which further added fixed effect slopes for demographic variables. Full model specifications can be found on OSF (osf.io/v9fst).

\begin{figure}[!t]
 \centering % avoid the use of \begin{center}...\end{center} and use \centering instead (more compact)
 \includegraphics[width=0.9\columnwidth]{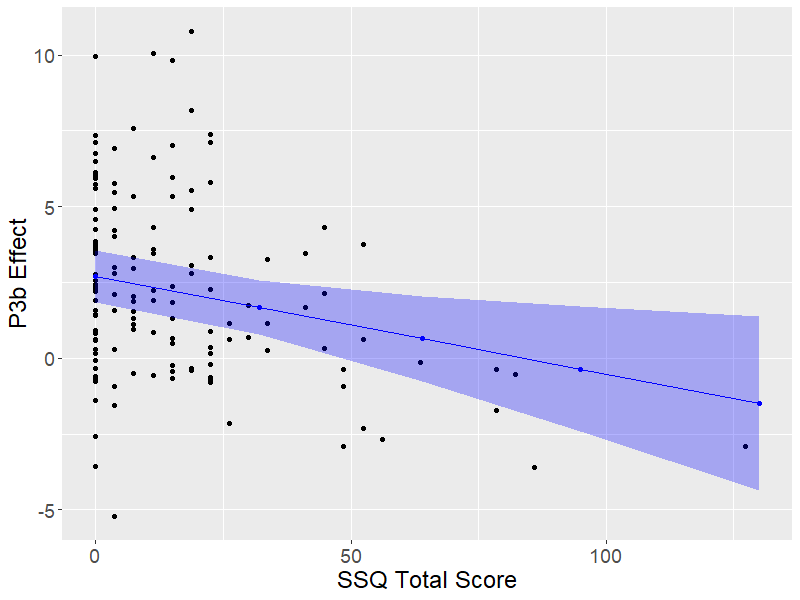}
 \vspace{-2mm}
 \caption{Model parameterized relationship between sickness severity and attention in full model with 95\% confidence bands.}
 \label{fig:HLM_Model}
 \vspace{-3mm}
\end{figure}

Likelihood ratio tests were used to compare model fits. Sickness severity on its own (the "sickness only" model) predicted P3b effect mean amplitudes significantly better than the null model, $\chi^2(1, \textit{N} = 29) = 15.98$, $\textit{p} < .001$, revealing that increasing sickness severity measured via SSQ TS at the end of each block accurately predicts smaller block mean P3b effects without the need for block labels. Congruent with the equivalent ANOVA analysis in Section~\ref{P3b_Effects_Block}, the condition only model also outperformed the null model, $\chi^2(4, \textit{N} = 29) = 24.33$, $\textit{p} < .001$. Critically, the full model which included sickness severity as an additional predictor beyond condition further outperformed the condition only model, $\chi^2(1, \textit{N} = 29) = 6.90$, $\textit{p} = .009$, and the sickness only model, $\chi^2(4, \textit{N} = 29) = 15.25$, $\textit{p} = .004$. The $\beta$ estimates for the full model predictors are plotted in Fig.~\ref{fig:Beta_points}, which represent HLM predictor regression coefficients where the magnitude indicates the strength of the relationship, and the sign represents the direction of the relationship, relative to the intercept (the "no sickness" condition in the case of block types and an SSQ TS of 0 in the case of SSQ data). Blocks that had more sickening paths (or that came later; note that the individual contributions of block type and exposure duration are not possible to separate due to the fixed block order of the design) were associated with greater attentional impacts and smaller differences between the oddball and standard P3b mean amplitudes (P3b effects), and likewise, higher SSQ TS was associated with smaller P3b effects, beyond the variance captured by block type. The full model's predicted relationship between sickness severity (as indexed by SSQ TS) and attentional reserve (as indexed by the P3b effect) is plotted in Fig.~\ref{fig:HLM_Model}. The effect of sickness severity in the full model was modest but statistically significant, $\beta = -0.032$, $95\%$ CI = [-0.056, -0.008]. Adding demographic predictors did not further improve model performance, $\chi^2(3, \textit{N} = 29) = 4.36$, $\textit{p} = .23$, and the predictive value of the available demographic information appeared negligible in our case (all $\beta$ $95\%$ CIs included 0). The MLM analysis thus reveals an important impact of sickness severity on attentional resources, as SSQ total scores were not only associated with P3b effects on their own, but significantly improved model performance over and above that which was provided by block type information.

\subsubsection{Behavior}

Having established a relationship between sickness severity and attentional resources, we were further interested in understanding how each might contribute to behavior, which in our case was most explicitly measured in the form of count errors of the number of oddballs presented in each block. If sickness reduces attention, then we should expect to see a greater number of count errors as participants get sicker. Zero-inflated poisson (ZIP) regression is a popular technique for modeling data of this nature \cite{lambert1992zero} and the R package `glmmTMB' \cite{Brooks_glmmTMB} used here is particularly convenient for adding multi-level structure to ZIP models. A model with only an SSQ TS predictor ($\beta = 0.015$, $95\%$ CI = [0.008, 0.022]) significantly outperformed its relative null model, $\chi^2(1, \textit{N} = 29) = 15.09$, $\textit{p} < .001$. The same was true in the case of a model with only a P3b predictor ($\beta = -0.135$, $95\%$ CI = [-0.204, -0.066]), $\chi^2(1, \textit{N} = 29) = 14.42$, $\textit{p} < .001$), indicating that sickness severity, attentional allocation, or some combination of the two confer effects on participant behavior. 

\begin{figure}[!t]
 \centering % avoid the use of \begin{center}...\end{center} and use \centering instead (more compact)
 \includegraphics[width=\columnwidth]{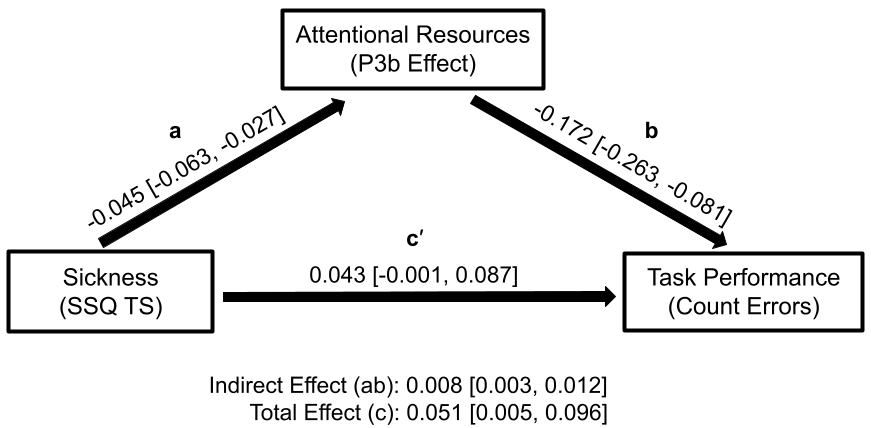}
 \vspace{-5mm}
 \caption{Mediation model with regression coefficients and 95\% CIs.}
 \label{fig:mediation_model}
 \vspace{-5mm}
\end{figure}

\subsubsection{Mediation Analysis}

The highly correlated nature of sickness severity and attention renders using them both as predictors in the same model of count errors problematic. This problem is due to multi-collinearity, a well-known issue in regression models where it is impossible to separate individual contributions of highly correlated variables. We reasoned, however, that increasing sickness would lead to reduced attentional resources, which would in turn lead to a greater number of count errors. We turned to mediation analysis to explicitly test this idea. 

A mediation model was fit using the structural equation modeling package `lavaan' in R \cite{Rosseel_lavaan}, with sickness as a predictor variable, via SSQ TS, attentional resources as a mediator variable, via the P3b effect, and behavioral performance as the outcome variable, via oddball count errors. A graphic of the mediation model is presented in Fig.~\ref{fig:mediation_model}. The model supported our rationale; as sickness increased, attentional resources decreased, and as attentional resources decreased, count errors became more numerous.

\section{Discussion}

This study sought to investigate if VR sickness has an impact on attention, and if those changes in attention can be successfully measured with EEG during HMD use. Though many previous researchers have investigated causes for VR sickness and methods to measure its symptoms \cite{Chang_Kim_Yoo_2020}, the cognitive and behavioral impacts of VR sickness are less well understood. Vection induced outside of VR has been shown to reduce the amplitude of the P3b \cite{Wei_Okazaki_So_Chu_Kitajo_2019} and affect task performance \cite{Strozak_etal_2016}, and a reduction in single-task P3b amplitudes as well as an increase in the amount of errors after exposure to a sickening stimuli in VR has been found \cite{Wu_Zhou_Li_Kong_Xiao_2020}. Although changes in the P3b have been found in a single-task paradigm using VR sickness-inducing oddballs \cite{Ahn_Park_Jeon_Lee_Kim_Hong_2020}, to our knowledge, changes in attention have not been measured while varying levels of VR sickness were induced with a dual-task paradigm. Thus, we sought to elucidate how VR sickness severity affected attention to a secondary task while in VR. We believed that increasing amounts of VR sickness would involuntarily redirect greater amounts of attention towards unpleasant internal bodily sensations as they developed, thus reducing the amount of attentional reserve available for completing an oddball counting task. Our results provide support for this theory. We found a statistically significant reduction in the amount of attention between the no sickness, low sickness, and high sickness conditions.

Since the observed attenuation in P3b effect mean amplitudes between blocks was able only to indicate the extent, but not the cause, of the attentional impacts, it was important to test if these effects were related to VR sickness specifically. First, we examined whether simply using VR led to a significant reduction in attention. We found no significant difference in P3b effect mean amplitudes between the first single-task block without VR and the no sickness dual-task block in VR. Thus, the interesting nature of immersion in VR by itself did not appear to result in discernible differences in the ability to attend to the oddball task. Motion in VR, however, did result in reliably attenuated P3b amplitudes. These effects were not observed immediately but emerged as a difference between the first low sickness condition block and the second high sickness condition block. As we used a fixed order of blocks to increase the power to detect an effect, the accumulation of sickness symptoms across multiple blocks could have resulted in these differences. These results are congruous with the finding that motion in VR had an impact on cognitive load only for subjects who were susceptible to VR sickness \cite{Ahn_Park_Jeon_Lee_Kim_Hong_2020}. We did not find a correlation between the change in severity of sickness across the study with the change in P3b effect mean amplitudes from the first to the last single-task blocks; however, carryover effects of lingering sickness on attention may still explain the significant difference in P3b amplitudes between the single-task block at the start and the single-task block at the end of the experiment, which is consistent with previous research \cite{Wu_Zhou_Li_Kong_Xiao_2020}. 

There was significant individual variation in the amount and timing of sickness symptoms. Several participants experienced greater sickness severity during the first high sickness block than the second high sickness block, so it is possible that adaptation to the sensory conflict developed in competition with or instead of accumulation effects. It was therefore necessary to examine whether there was a relationship between changes in P3b amplitudes, as an index of attentional reserve, and the intensity of VR sickness symptoms experienced independent of when in the study the sickness occurred. We theorized that the blocks and conditions which we categorized as low or high sickness might not correspond to when subjects actually experienced high and low amounts of VR sickness, and thus artificially grouping participants in this way might mask the effects. Using MLM allowed us to account for this individual variability and outliers within the participant data structure. Consistent with our expectation that increased sickness would decrease attentional resources available for the oddball task, we observed that P3b effect mean amplitudes showed significant differences by block, and furthermore that individual sickness scores explained more of the variance in P3b effects than the difference between blocks alone. That the sickness intensity regardless of block predicted the difference in attention, as measured in ERPs, provides evidence that the changes in attention were in fact due to sickness. Further support for VR sickness as a source for the depreciation in attention was uncovered in the relationship between the intensity of sickness symptoms and the impairment in task performance in counting the oddball tones that occurred during each block, similar to previous work that found a negative correlation between VR sickness and task performance \cite{Sepich_Jasper_Fieffer_Gilbert_Dorneich_Kelly_2022}. The reduction in P3b effect mean amplitudes mediated the increase in SSQ scores and the increase in the number of errors. Thus, as VR sickness escalated, attentional reserve was depleted and the number of errors in the oddball counting task increased.

In this study, we investigated VR sickness as a constraint on attentional resources, thus reducing focus on the completion of the primary and secondary tasks. With this study design, it was not possible to determine how much attention was captured by each of the three competing constraints, but the combination did result in a significant reduction in P3b amplitudes. The primary task was not designed to be especially difficult, and our measure of performance on that task was set up just to rule out non-compliance. As the museum tours were repeated and the artwork quiz was given at the end of the study, we expected good performance on the quiz even if participants got too sick to pay close attention to the pieces of art on each tour. Nevertheless, performance on the secondary task, which was measured during each block, was impaired as attention decreased, suggesting participants were indeed prioritizing the primary task as instructed. If the combination of tasks and sickness did not sufficiently deplete attentional resources, we would not have detected a significant reduction in P3b amplitudes. Therefore, our finding that SSQ scores are associated with P3b amplitudes even with less demanding primary and secondary tasks perhaps attests to the power of VR sickness to capture significant attentional resources, as well as the power of this approach to capture even subtle attentional shifts when other attentional demands are not as intense. Physiological arousal is theorized to be a determining factor in processing capacity \cite{Kahneman_1973, Polich_2007}, which may play a role in the case of VR sickness.

There are many internal and external stimuli unrelated to the experiment that can attract attention, so it can be challenging to be sure that observed attentional fluctuations are primarily the result of experimental manipulations. Therefore, dual-task paradigms have been particularly illuminating in tracking changes in attentional reserve \cite{Vidulich_Tsang_2012}. ERP dual-task studies have explored cognitive impacts during physical locomotion \cite{Nenna_Do_Protzak_Gramann_2021, Ladouce_Donaldson_Dudchenko_Ietswaart_2019, Robles_Kuziek_Wlasitz_Bartlett_Hurd_Mathewson_2021}, and dual-task auditory paradigms in VR have also been used to investigate presence \cite{Kober_Neuper_2012, Grassini_Laumann_Thorp_Topranin_2021}. In early work with a dual-task auditory paradigm, Kramer and colleagues \cite{Kramer_Sirevaag_Braune_1987} compared P3b amplitudes during missions on a flight simulator. In their study, the difficulty in the primary task was increased by changing aspects of the flight missions, whereas in our study the primary task remained the same and the increase in difficulty was related to the development of VR sickness symptoms. They found decreases in the secondary task P3b amplitudes during increases in task demands, analogous to our finding of decreases in P3b amplitudes during increases in VR sickness. They also found a correlation between reductions in P3b amplitudes and primary task performance, which is comparable to the inverse relationship between P3b amplitudes and errors on the secondary task shown in our mediation analysis. More recently, ERP studies during virtual driving simulations have investigated attentional effects \cite{Fang_Zhang_Zhang_Fang_2020} related to mind-wandering \cite{Baldwin_Roberts_Barragan_Lee_Lerner_Higgins_2017}, distracted driving \cite{Banz_Wu_Camenga_Mayes_Crowley_Vaca_2020}, and alcohol use \cite{Wester_Verster_Volkerts_Böcker_Kenemans_2010}. Using a single-task paradigm, Ahn et al. \cite{Ahn_Park_Jeon_Lee_Kim_Hong_2020} found larger P3b amplitudes in a subgroup of subjects with higher susceptibility to VR sickness based on SSQ scores. The researchers speculated that this was due to increased recruitment of cognitive resources during sensory conflict. As increases in cognitive load are associated with diminished attentional reserve \cite{Jaquess_Gentili_Lo_Oh_Zhang_Rietschel_Miller_Tan_Hatfield_2017}, their results provide support for our hypothesis. If sensory conflict and VR sickness from vection in VR increase cognitive load and divert attentional resources away from the secondary task during a dual-task paradigm, these changes in attentional reserve would be reflected in reduced amplitudes of the P3b, as we found. Thus, by building upon previous work on attentional impacts with ERP measures %\cite{Ghani_Signal_Niazi_Taylor_2020, Vidulich_Tsang_2012} 
and testing with a dual-task paradigm in an HMD, our study provides new evidence of the deleterious consequences of VR sickness.

The implication of our results are important to consider for any developers that want their users to learn skills applicable to real-world scenarios while in VR. If participants are not able to pay full attention or their performance is impaired, the benefits conferred by an HMD may be diminished. For example, immersive applications where attention and task performance are critical, like VR training and simulations, may need to include factors associated with higher sickness, such as controller-based locomotion or longer exposure time \cite{Saredakis_Szpak_Birckhead_Keage_Rizzo_Loetscher_2020, Caserman_Garcia-Agundez_Gamez-Zerban_Gobel_2021}. Researchers might want to consider the confound of VR sickness on any measures they collect in VR, and developers could apply our results by considering trade-offs when creating content that could increase VR sickness. Researchers could also adapt this method to address these concerns. When the design is conducive to a secondary task that requires attention, a dual-task paradigm could be used to explore how much of a sickness and performance impact certain choices have by comparing changes in P3b amplitudes across conditions with different elements. Changes in cognitive load between conditions can impact attentional reserve, so this must be taken into consideration. The P3b is robust enough to use in real-time \cite{Allison_Kubler_Jin_2020}, so this method provides the benefit of tracking changes in attentional reserve as they occur, without requiring that subjects are aware of or continuously report on their sickness during VR exposure.

It can be convenient to separate the modality of the second task oddball paradigm from that of the primary task, as we did here, to make the tasks easier to distinguish and perform simultaneously. For our study, the P3b auditory oddball paradigm provided certainty over the timing of the ERP stimulus, which can be more difficult to track with a visual task in an HMD. However, previous studies have used simultaneous visual cues \cite{Strozak_etal_2016, Nenna_Do_Protzak_Gramann_2021}. Gaming content, which can elicit high SSQ scores and dropouts \cite{Saredakis_Szpak_Birckhead_Keage_Rizzo_Loetscher_2020, Caserman_Garcia-Agundez_Gamez-Zerban_Gobel_2021}, could be modified to include a visual paradigm. There is some evidence that as VR sickness increases, visual exploration of the environment decreases \cite{lee_kim_kim_lee_2021}, but targets could appear in the virtual environment which subjects have to mentally count or press a button in response to as a secondary task while prioritizing completion of some other primary task (see \cite{Vidulich_Tsang_2012, Ghani_Signal_Niazi_Taylor_2020} and \cite{Polich_2007, Kok_2001, Kok_1997, Verleger_2020} for discussion on considerations when designing these tasks). A few groups have assessed attention and workload with a method that does not require subjects to attend to the secondary task using novel task-irrelevant probes to compare a different ERP response (the P3a)  \cite{Miller_Rietschel_McDonald_Hatfield_2011, Jaquess_Gentili_Lo_Oh_Zhang_Rietschel_Miller_Tan_Hatfield_2017, Allison_Polich_2008}. We targeted the P3b instead, first because frontal electrodes (where P3a amplitudes are larger) may have greater interference from eye and VR headset-related artifacts, and also because ERP responses to unanticipated stimuli may be harder to elicit if the participants are so engaged in the primary task that their attention is not captured by the probes, such that smaller ERP responses can arise both from engagement and disengagement.

\section{Limitations and Future Work}

One potential limitation of the study was an increase in the prevalence of artifacts. Muscle activity, eye blinks, and electrical noise can all impede data collection \cite{Mumtaz_Rasheed_Irfan_2021}. It was therefore important to confirm that our data were sufficiently clean. Similar to previous studies \cite{Aksoy_Ufodiama_Bateson_Martin_Asghar_2021, Harjunen_Ahmed_Jacucci_Ravaja_Spape_2017}, we did not find substantial interference from the HMD. We used three different preprocessing pipelines and compared ERP measurements. Between the pipelines, preprocessing did not strongly impact P3b effect mean amplitude differences. Further details regarding the pipelines and the results from the comparison are available on OSF (osf.io/v9fst). Another possible limitation is that our inclusion criteria necessarily restricted the range of VR sickness participants were likely to experience. Our IRB approval was contingent on our notifying participants before signing up that the study involved VR content that could cause nausea, and people that were very sensitive to motion sickness were discouraged from signing up for the study. Thus, the sample of participants was self-selected for individuals less prone to extreme sickness reactions. Third, the sample size was small due to a limited time frame in which we had access to EEG facilities. However, our sample size is comparable to many other EEG studies, and we expected that the P3b effect would be quite large based on extensive previous research on this component \cite{Polich_2007}. Finally, although we had an a priori hypothesis about the direction of the change in the P3b under conditions of increasing VR sickness, which the study was designed to test, the comparisons of preprocessing pipelines and analysis parameters were not pre-registered and thus exploratory. We will run a confirmatory follow-up study to provide further support our results.

\section{Conclusion}

In this study, we sought to address an important gap in the literature regarding the impact of VR sickness on attention to tasks performed in an HMD. We found a significant difference in participants' attention to a secondary oddball task, as measured with the amplitude of the P3b, between the VR sickness conditions, and furthermore that the SSQ scores are reliably linked to the effect mean amplitude of the P3b and the number of errors on the oddball counting task. Thus, we provide preliminary evidence that VR sickness has a deleterious effect on attention, which in turn leads to impairment in task performance. We provide subject data through Harvard Dataverse as well a preprocessing script for EEGLAB which can be used by other researchers. More generally, the dual-task oddball paradigm described here could be used to compare attentional impacts with other VR applications.

\acknowledgments{
The authors thank Ryan Hubbard, Melinh Lai, Melissa Troyer, and members of the CAB Lab for feedback on the project, as well as Michael N. Mimnaugh and Başak Sakçak for feedback on the manuscript. This work was supported by a European Research Council Advanced
Grant (ERC AdG, ILLUSIVE: Foundations of Perception Engineering, 101020977), the Academy of Finland (project PERCEPT 322637), Business Finland (project HUMOR 3656/31/2019), CONACYT (project 745), and scholarships from Google and the Finnish Foundation for Technology Promotion.

}

\bibliographystyle{abbrv-doi-hyperref}

\bibliography{references}

\end{document}